\documentclass{article}\usepackage{graphicx}
\usepackage{epsfig}
\usepackage{amsfonts}
 \usepackage{amssymb}
\usepackage{amsmath}
\usepackage{slashed}

\usepackage{hyperref}
\usepackage{graphicx}       
\usepackage{dcolumn}        
\usepackage{bm}             
\usepackage{amssymb}
\usepackage{amstext}
\usepackage{tensor}

\date{\today}

\newcommand{\insertplot}[5]{\begin{figure}
 \hfill\hbox to 0.05in{\vbox to #5in{\vfill
 \inputplot{#1}{#4}{#5}}\hfill}
 \hfill\vspace{-.1in}
 \caption{#2}\label{#3}
 \end{figure}}
 \newcommand{\inputplot}[3]{
 \special{ps: plotfile #1}
}
\newcounter{fig}

\newcommand{\ee}{\end{equation}}
\newcommand{\eea}{\end{eqnarray}}
\newcommand{\bea}{\begin{eqnarray}}

\newcommand{\beq}{\begin{equation}}
\newcommand{\eeq}{\end{equation}}

\begin{document}

\title{\Large{\bf Asymptotically flat scalar, Dirac and Proca stars: \\ discrete $vs.$ continuous families of solutions}}

\author{
{\large Carlos A. R. Herdeiro,}\footnote{herdeiro@ua.pt} \ 
{\large Alexandre M. Pombo}\footnote{ pomboalexandre@ua.pt} \ 
and
{\large Eugen Radu}\footnote{eugenradu@ua.pt}
\\ 
\\
{\small Departamento de F\'\i sica da Universidade de Aveiro and} \\
{\small  Center for Research and Development in Mathematics and Applications (CIDMA)} \\ 
{\small Campus de Santiago, 3810-183 Aveiro, Portugal}
}
\date{August 2017}
\maketitle

\begin{abstract}   
The existence of localized, approximately stationary, lumps of the classical gravitational and electromagnetic field -- \textit{geons} -- was conjectured more than half a century ago. If one insists on \textit{exact} stationarity, topologically trivial configurations in electro-vacuum are ruled out by no-go theorems for solitons. But stationary, asymptotically flat geons found a realization in scalar-vacuum, where everywhere non-singular, localized field lumps exist, known as (scalar) \textit{boson stars}. Similar geons have subsequently been found in Einstein-Dirac theory and, more recently, in Einstein-Proca theory. We identify the common conditions that allow these solutions, which may also exist for other spin fields. Moreover, we present a comparison of spherically symmetric geons for the spin $0,1/2$ and $1$, emphasising the mathematical similarities and clarifying the physical differences, particularly between the bosonic and fermonic cases. We clarify that for the fermionic case, Pauli's exclusion principle prevents a continuous family of solutions for a fixed field mass; rather only a discrete set exists, in contrast with the bosonic case.
\end{abstract}

\section{Introduction and overview} 
	
In 1955~\cite{Wheeler:1955zz}, John Wheeler investigated the existence, within general relativity (GR) coupled to classical electromagnetism, of \textit{``classical, singularity free, exemplar of the  ``bodies" of classical physics"}. He named such (material) source-free entities \textit{geons}, after ``{\bf g}ravitational {\bf e}lectr{\bf o}magnetic e{\bf n}titie{\bf s}" and wrote:

\bigskip

\textit{``The simplest variety is most easily visualized as a standing electromagnetic wave, or beam of light, bent into a closed circular toroid of high energy concentration. It is held in this form by the gravitational attraction of the mass associated with the field energy itself. It is a self-consistent solution of the problem of coupled electromagnetic and gravitational fields (...)"} [p. 512]

\bigskip

Wheeler, however, could not provide a complete solution of the Einstein-Maxwell equations describing  geons, which motivated subsequent attempts at obtaining not only electromagnetic ($e.g.$~\cite{Ernst:1957zz,Melvin:1963qx}), but also neutrino~\cite{Brill:1957fx} and purely gravitational geons~\cite{Brill:1964zz}.  
In asymptotically flat
 (electro-)vacuum, such discussion is not fully settled. The dominating view is that no topologically trivial, \textit{stable} geons exist (see $e.g.$~\cite{Anderson:1996pu,Perry:1998hh}).

The original proposal of geons does not require \textit{precise} stationarity. In fact, (electro-)vacuum stationary, asymptotically flat, everywhere regular configurations are ruled out by classical theorems in GR~\cite{Einstein,Lichnerowicz,Heusler:1996ft}. But a realization of stationary geons was discovered in the Einstein-Klein-Gordon system~\cite{Kaup:1968zz,Ruffini:1969qy}. These topologically trivial, localized gravitating scalar solitons, are known as  \textit{boson stars} (BSs)~\cite{Schunck:2003kk,Liebling:2012fv}. 

BSs have a stable branch (against linear perturbations)~\cite{Gleiser:1988ih,Lee:1988av}. Their existence is based on three key properties:
\begin{description}
\item[(i)] the field is composed of standing waves oscillating with some frequency; 
\item[(ii)] there is a confining mechanism for the field;
\item[(iii)]  the energy-momentum tensor is invariant under the timelike Killing vector field.
 \end{description}
Property ${\bf (i)}$ realizes Wheeler's vision: the energy lump is made of self-gravitating standing waves. The oscillation originates an effective pressure that counter-acts the tendency for gravitational collapse, within GR\footnote{For asymptotically flat geons in alternative theories of gravity, see, $e.g$,~\cite{Olmo:2017qab,Afonso:2017aci}.} and without resorting to energy conditions violating matter. Mathematically, the explicit harmonic time dependence in the field, 
evades virial-type arguments that rule out the absence of solitons \cite{Heusler:1996ft}; such arguments  are gravitational extensions of Derrick's theorem in field theory~\cite{Derrick,Hobart}.


 In Wheeler's vision the  standing waves' self-gravity should be enough to create a (sufficiently) stable energy lump. But the limited success with electro-vacuum geons indicates this is insufficient; a confining mechanism is necessary -- property ${\bf (ii)}$. For BSs, this mechanism is (typically) the field's mass $\mu$, creating a potential barrier at spatial infinity and gravitationally binding waves with frequency $w<\mu$. 
 
 
 Still, these two properties do not suffice to create a \textit{stable} energy lump. Field oscillations generate, via the non-linearities of GR, higher frequency harmonics, which leak towards infinity overcoming the gravitational potential well (and the mass potential barrier). This is explicitly seen in  \textit{oscillatons}~\cite{Seidel:1991zh}, real scalar field configurations with a fundamental oscillation frequency. Property ${\bf (iii)}$ prevents this. It is realized, for BSs, by having two standing waves with the same frequency but opposite phases  - Fig. 1 -, canceling out all dynamics at the level of the energy-momentum tensor.

 \begin{figure}[h!]
\begin{center}
\includegraphics[width=0.242\textwidth]{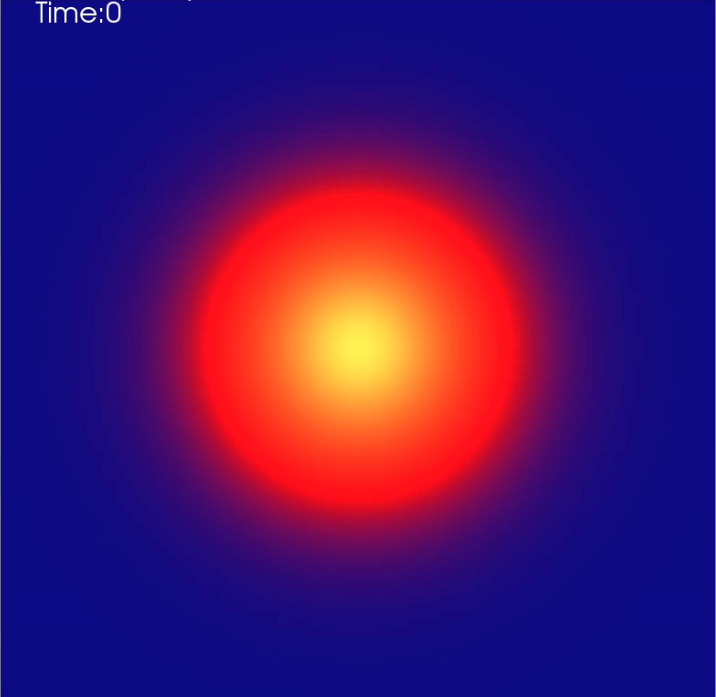}\ 
\includegraphics[width=0.242\textwidth]{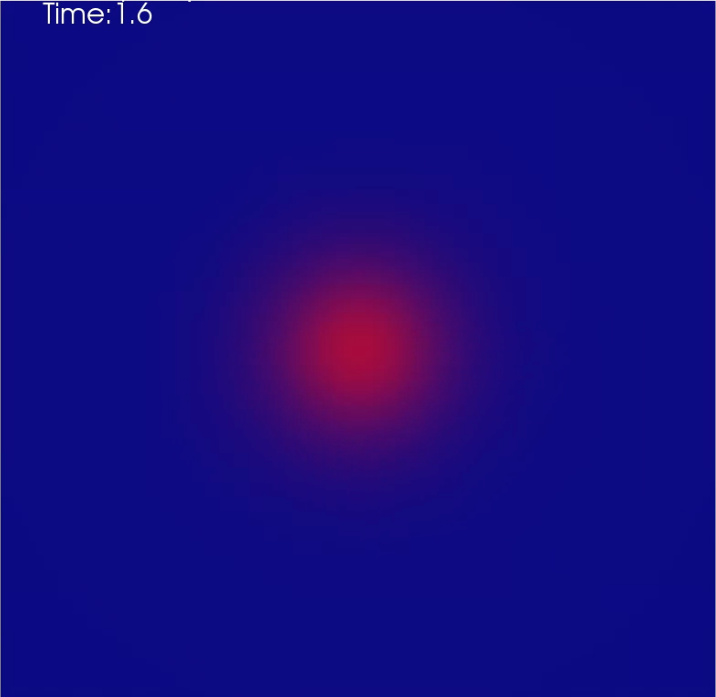}
\includegraphics[width=0.242\textwidth]{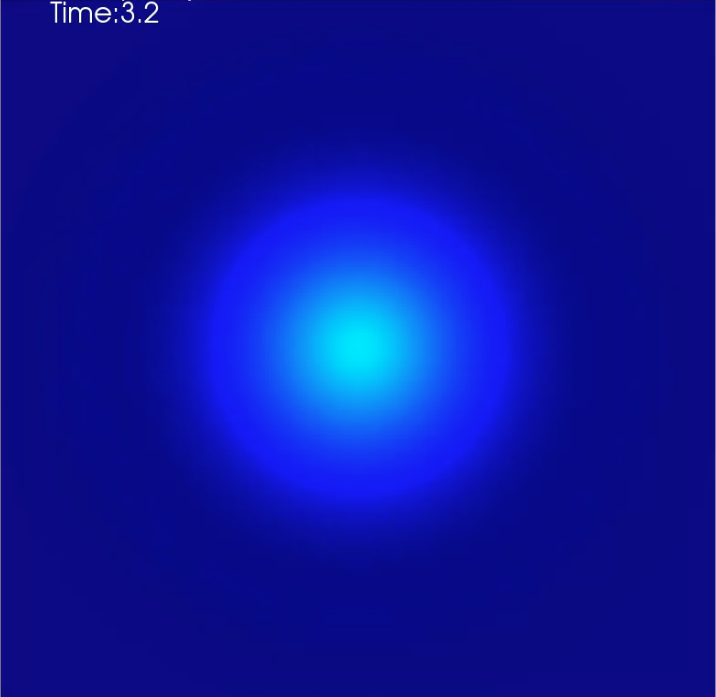}
\includegraphics[width=0.242\textwidth]{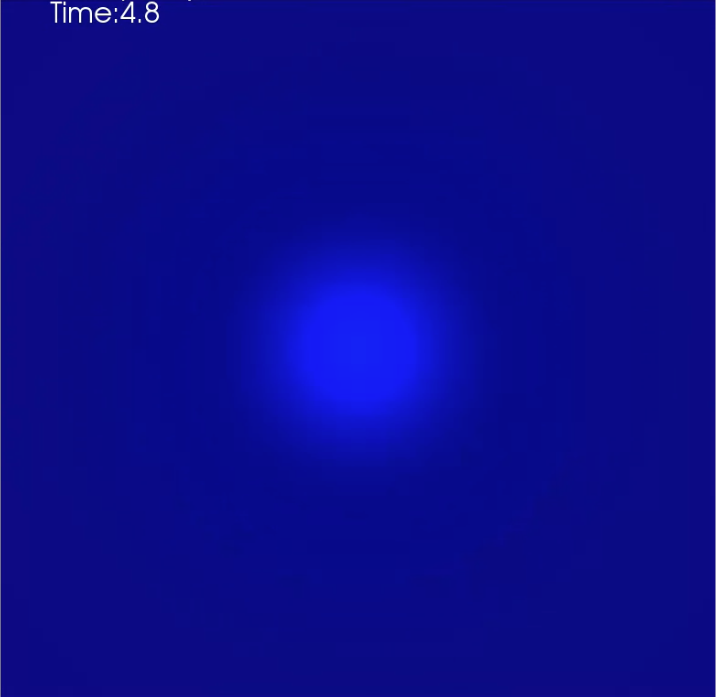} \\ \vspace{0.1cm}
\includegraphics[width=0.242\textwidth]{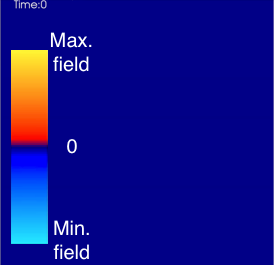}\ 
\includegraphics[width=0.242\textwidth]{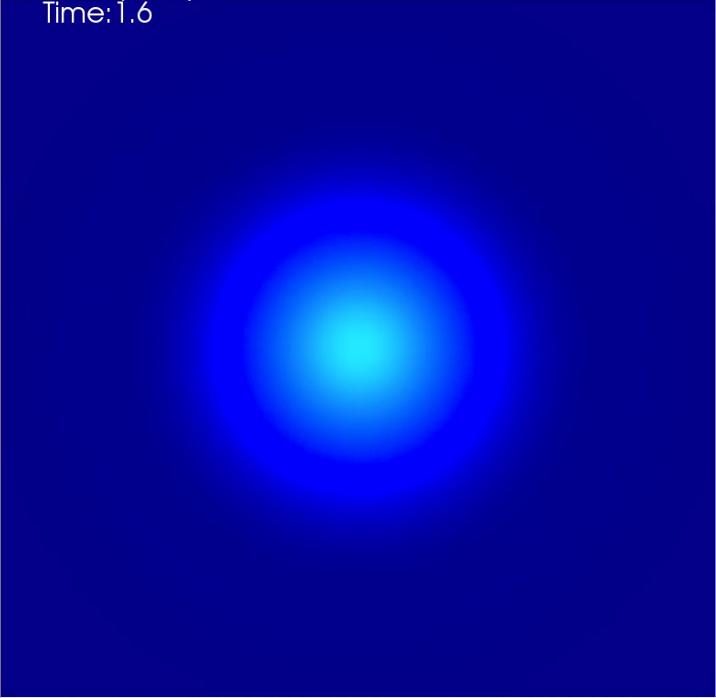}
\includegraphics[width=0.242\textwidth]{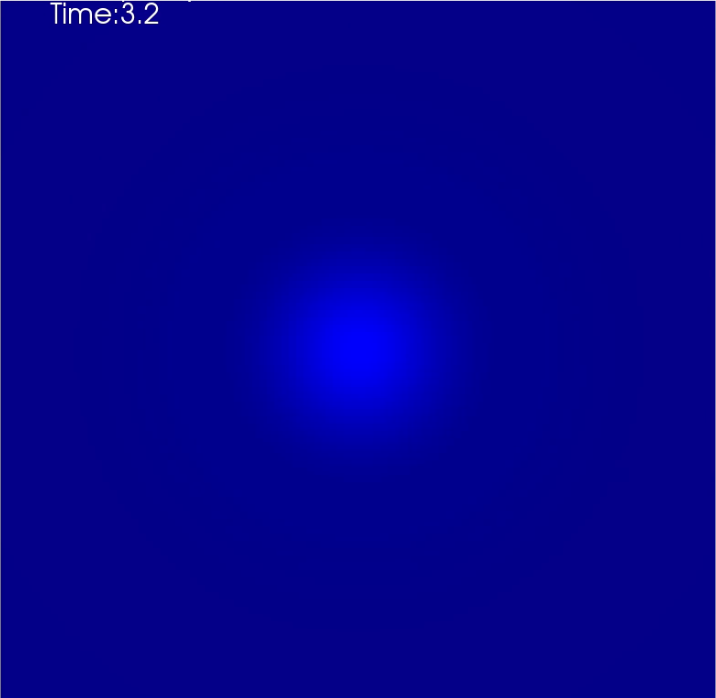}
\includegraphics[width=0.242\textwidth]{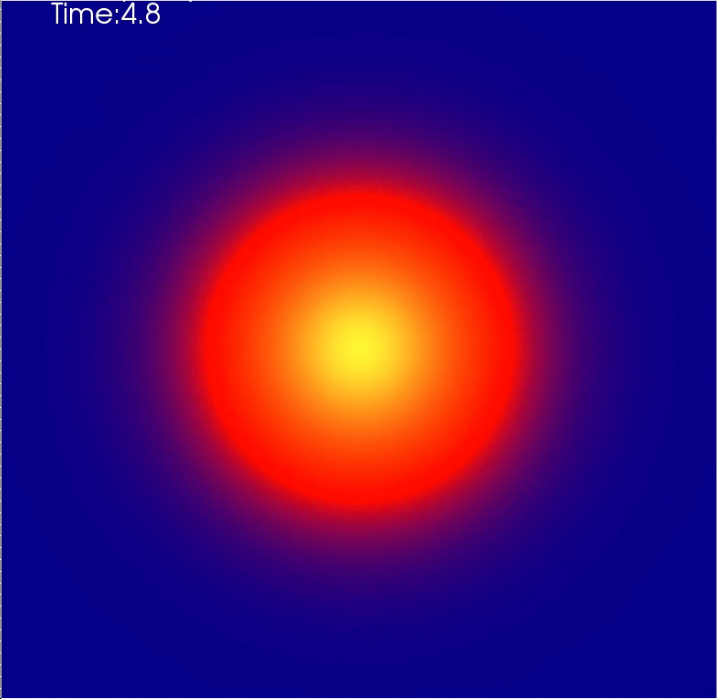}
\caption{\small{Four snapshots at roughly (but not precisely) intervals of one quarter of the period of the two scalar field modes, corresponding to the real and imaginary parts of the complex scalar field $\Phi$, $cf.$ Sec.~\ref{sec2}, in a numerical relativity fully non-linear time evolution of a spherical, stable BS~\cite{zilhao}.} In the first snapshot the top mode is at the maximum and the bottom one vanishes. The energy-momentum tensor is time independent. Axisymmetric rotating configurations also exist~\cite{Schunck:2003kk}, which, as Wheeler imagined, are toroidal energy distributions~\cite{Schunck:1996he,Yoshida:1997qf}.}
\label{fig1}
\end{center}
\end{figure}  

\begin{figure}[h!]
\begin{center}
\includegraphics[width=0.49\textwidth]{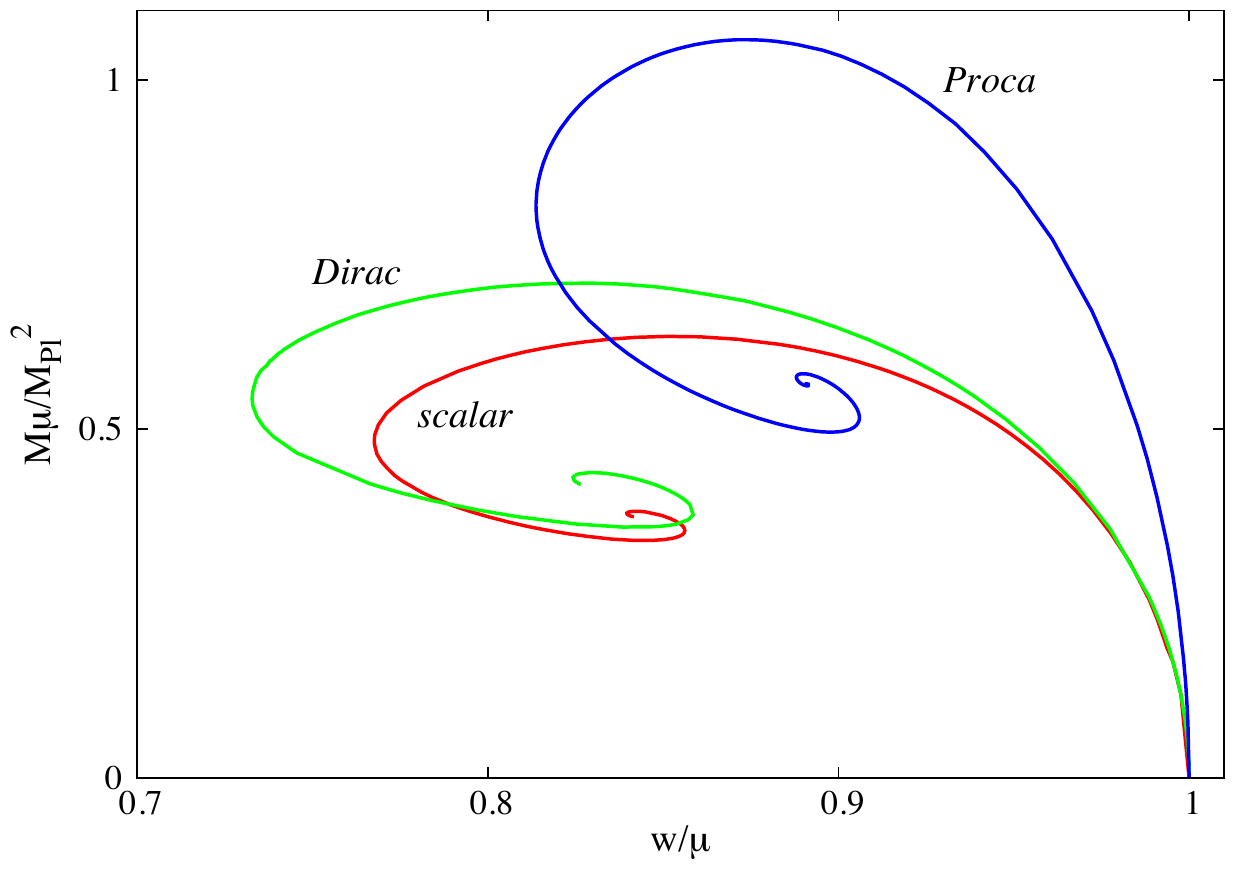}
\includegraphics[width=0.49\textwidth]{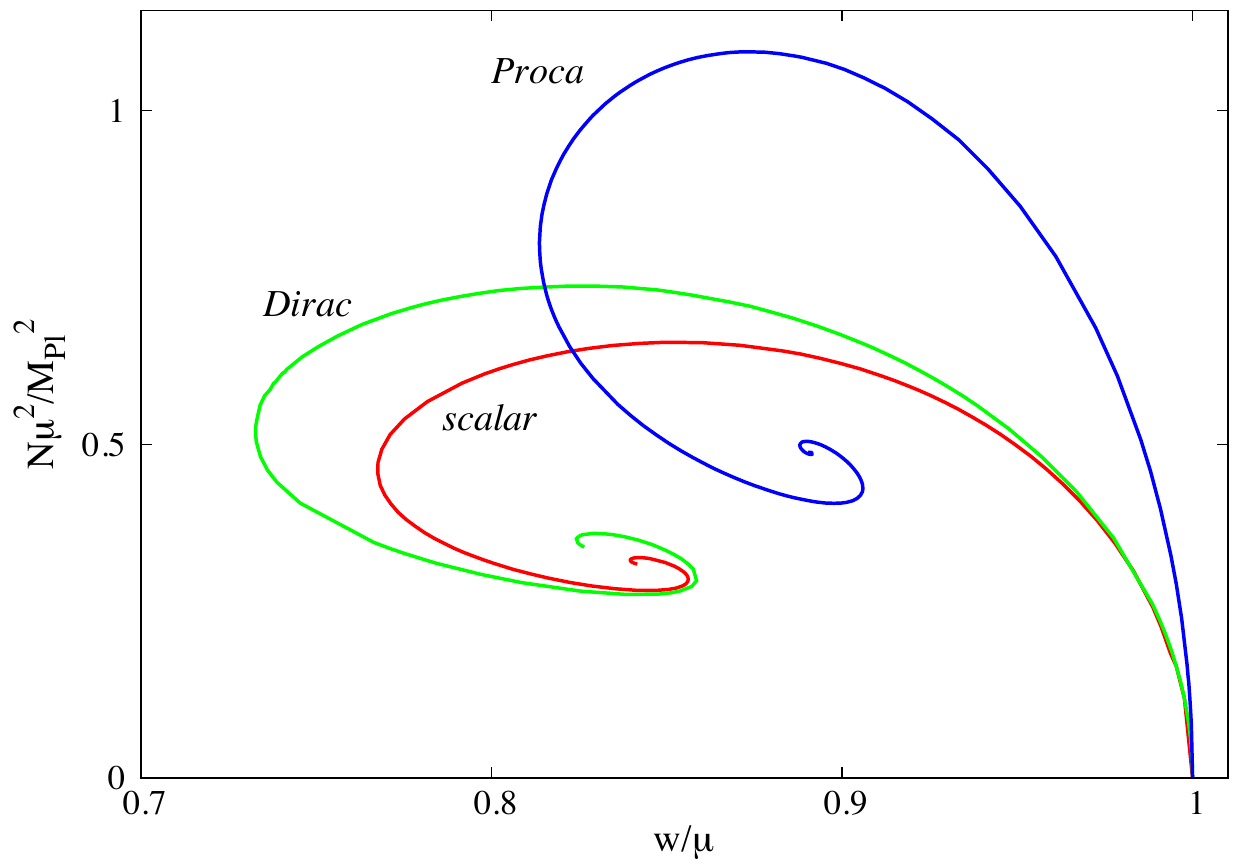}  
\caption{\small{The ADM mass $M$ (left) and the 
particle number $N$ (right) $vs.$ field frequency for the minimal scalar (red line), vector (blue line) and spinor (green line) models. The details explaining the construction of these plots are given in Sec.~\ref{sec4}.}}
\label{fig2}
\end{center}
\end{figure}

One way in which BSs depart significantly from Wheeler's vision is that they are made of waves with a \textit{single} frequency.\footnote{
According to Wheeler [1], when attempting to find a spherically symmetric geon, \textit{``The different elementary disturbances must have different frequencies. If all had the same frequency, they would add coherently to form a single mode of distribution of electromagnetic field strength. But there is no such thing as a nonzero source-free spherically symmetrical electromagnetic field disturbance."} [p. 518]}
BSs are composed of many coherent modes. Depending on the frequency and on the particular model 
there is a discrete set of BS solutions with that frequency, corresponding to a fundamental state and a set of excited states.  
 In Fig.~\ref{fig2}  (left, red solid line)
 we show the ADM mass  of the fundamental states of BSs ($cf.$ the model in Sec.~\ref{sec2}), $M$, $vs.$ the scalar field frequency, $w$,  in units of the scalar field mass $\mu$ and Planck mass $M_{Pl}$. 
Observe that: $1)$ BSs require a minimal frequency 
and they only exist up to a maximal mass, $cf.$ Property ${\bf (i)}$. 
Intuitively, only fast enough oscillations, but not too energetic, prevent gravitational 
collapse. 
$2)$ solutions only exist for $w<\mu$, $cf.$ Property ${\bf (ii)}$.

The maximal ADM mass, depends solely on the scalar field mass $\mu$ and is of the form~\cite{Kaup:1968zz} 
\begin{equation}
M^{\rm max}=\alpha_s \times 10^{-19}M_\odot\left(\frac{\rm GeV}{\mu}\right) \ , \ \ {\rm  with}
 \ \ \alpha_0 \simeq 0.633 \ \ \ \ \ (s={\rm spin})\ .
\label{maxm}
\end{equation}
Thus, only ultra-light bosons, with a mass $\mu\lesssim 10^{-19}$ GeV, can source stellar mass BSs. These occur in a variety of  ``beyond the Standard Model'' scenarios, most notably in the string axiverse~\cite{Arvanitaki:2009fg}.

How can a single frequency state, with this ultra-light mass create a star-like object? Because it has a very large \textit{occupation} (or \textit{particle}) number, $N$, which can be estimated by computing the Noether charge associated to the $U(1)$ global symmetry of the scalar model, that rotates the two modes (see Sec.~2 below). 
Upon quantization, this becomes an integer. 
Along the (red) spiral line of Fig.~\ref{fig2} (right panel) the occupation number $N$ varies similarly to the ADM mass -  Fig.~\ref{fig2} (left panel).  
  At its maximum, taking (say) $\mu \sim 10^{-19}$ GeV, implies a large number the order of $N\sim 10^{76}$. 
Macroscopic BSs are therefore \textit{macroscopic quantum states}, indeed macroscopic Bose-Einstein condensates -- see $e.g.$~\cite{Suarez:2013iw} for deepening this perspective and its relation to dark matter models.

It turns out that Properties ${\bf (i)}$-${\bf (iii)}$ can be reproduced for other spin fields. 
Spin 1 geons, akin to BSs, were recently 
constructed \cite{Brito:2015pxa}, 
in GR minimally coupled to a complex Proca field (mass $\mu$), dubbed \textit{Proca stars}. They mimic closely the scalar case, including in the existence of a stable branch~\cite{Brito:2015pxa,Sanchis-Gual:2017bhw}. The corresponding (blue) line of solutions is displayed in Fig.~\ref{fig2}. 
More intriguingly, analogous solutions to the classical Dirac equation (mass $\mu$), 
minimally coupled to GR also exist \cite{Finster:1998ws}. 
We have reproduced them, hereafter dubbed \textit{Dirac stars}. The corresponding (green) line of solutions is displayed in Fig.~\ref{fig2}.

As  classical GR solutions, all these three cases are qualitatively similar:  in  their domains of existence 
-- Fig.~\ref{fig2}; 
in their maximal mass, always of the form~\eqref{maxm}, 
with $\alpha_{1/2} \simeq 0.709$ and $\alpha_1 \simeq 1.058$; 
and in the existence of a conserved Noether charge, 
always associated to a global $U(1)$ symmetry and providing a measure of the particle number $N$ 
in the single frequency state. 
This last point, however, also unveils a sharp distinction between the bosonic and fermionic cases. 
Even though we have treated the Dirac equation classically, 
its fermionic nature should be imposed at the level of the occupation number: 
at most a single particle, in accordance to Pauli's exclusion principle. 
So, how can we interpret the spiral in the right panel of Fig.~\ref{fig2}, 
which in the bosonic case corresponded 
to a sequence of solutions with different particle number?

\medskip
In this paper we perform a comparative analysis of these three different types of solitonic solutions of GR-matter systems, putting them together under a consistent set of notations and conventions, a task which, to our knowledge was not considered before in the literature.
The mathematical description of each of the three models is made in parallel to emphasise its similarities. But the physical interpretation is necessarily distinct, for fermions and bosons. In particular we make clear that whereas the bosonic configurations form \textit{a continuous sequence or family of solutions for a given field mass, fermionic solutions do not.}

This paper is organized as follows. In Sec.~\ref{sec2} we describe the basic equations of each of the three different models. 
Then, in Sec.~\ref{sec3} the ansatz, explicit field equations and some relations for physical quantities are provided for the three cases. We also comment on the units of the main physical quantities and some scaling symmetries of these equations, which are relevant for obtaining the solutions, in practice. In Sec.~\ref{sec4} we discuss the solutions in more detail, provide various physical results, which include, in particular, the ones summarized in Fig.~\ref{fig2}. We also clarify the physical interpretation of the sequences of fermionic solutions. Concluding remarks and some open questions are presented in Sec.~\ref{sec5}.

\section{The three models}
\label{sec2}

We consider Einstein's gravity minimally coupled with a spin-$s$ field, with $s=0,\frac{1}{2},1$. The corresponding action is (we use units with $c=1=\hbar$)
\begin{eqnarray}
\label{action}
\mathcal{S}=\int d^4 x \sqrt{-g} 
\left [
\frac{1}{16 \pi G} R
+
\mathcal{L}_{(s)}
\right] \ ,
\end{eqnarray}
where the three corresponding matter Lagrangians are:
\begin{eqnarray}
\label{LS}
&& \mathcal{L}_{(0)}= - g^{\alpha \beta}\bar \Phi_{, \, \alpha} \Phi_{, \, \beta} - \mu^2 \bar \Phi \Phi \ , \qquad \mathcal{L}_{(1)}= -\frac{1}{4}\mathcal{F}_{\alpha\beta}\bar{\mathcal{F}}^{\alpha\beta}
-\frac{1}{2}\mu^2\mathcal{A}_\alpha\bar{\mathcal{A}}^\alpha \ ,
\\
%
\label{LD}
&&
 \mathcal{L}_{(1/2)}^{[A]}=-i 
\left[
\frac{1}{2}
  \left( \{ \hat{\slashed D}  \overline{\Psi}^{[A]}  \} \Psi^{[A]} -
     \overline{\Psi}^{[A]} \hat{\slashed D}  \Psi^{[A]}
	\right)
+\mu \overline{\Psi}^{[A]}  \Psi^{[A]} 
\right]\ .
\end{eqnarray}
In this paper, the conventions for scalars are those in~\cite{Herdeiro:2015gia};  for fermions, we shall follow the same framework, including the definitions and  conventions, as in~\cite{Dolan:2015eua}. Finally, in the Proca field case, we shall use the notation and conventions in~\cite{Brito:2015pxa,Herdeiro:2016tmi}. In all cases, the overbar denotes complex conjugation. Moreover, 
\begin{itemize}
\item
$\Phi$ is a complex scalar field. Then the first Lagrangian in  (\ref{LS})
is equivalent to a model with two real scalar fields, $\Phi^R,\Phi^I$, under the relation
 $\Phi=\Phi^R+i\Phi^I$.
 \item
$\Psi$ is a Dirac 4-spinor, with four complex components, while the index $[A]$
corresponds to the number of copies of the Lagrangian.
For a spherically symmetric configuration,
one should consider (at least) two spinors, with the equal mass $\mu$.  
Indeed, a model with a single spinor necessarily possesses a nonzero angular momentum density and cannot be 
spherically symmetric. $\hat{\slashed D}\equiv \gamma^\mu \hat{D}_\mu$, where  $\gamma^\mu$ are the curved space gamma matrices and $\hat{D}_\mu=\partial_\mu+\Gamma_\mu$ is the spinor covariant derivative with $\Gamma_\mu$ being the spinor connection matrices~\cite{Dolan:2015eua}.
\item
$\mathcal{A}$ is a complex four potential, with the field strengths $\mathcal{F} =d\mathcal{A}$.
Again, the model can be described in terms of two real vector fields,  $\mathcal{A}=\mathcal{A}^R+i\mathcal{A}^I$.
\end{itemize}
In all cases, $\mu>0$ corresponds to the mass of the elementar quanta of the field(s).


Extremizing the action (\ref{action}) leads to a system of coupled Einstein-matter equations of motion. The Einstein equations read 
$G_{\alpha \beta}=8 \pi G T_{\alpha \beta}^{(s)}$, where $G_{\alpha \beta}$ is the Einstein tensor and the energy momentum tensor, $T_{\alpha \beta}^{(s)}$, is, for the scalar, Dirac and Proca cases, respectively,
\begin{eqnarray}
\label{TS}
&&
T_{\alpha \beta}^{(0)}=
\bar  \Phi_{ , \alpha}\Phi_{,\beta}
+\bar \Phi_{,\beta}\Phi_{,\alpha} 
-g_{\alpha \beta}  \left[ \frac{1}{2} g^{\gamma \delta} 
 ( \bar \Phi_{,\gamma}\Phi_{,\delta}+
\bar \Phi_{,\delta}\Phi_{,\gamma} )+\mu^2 \bar \Phi\Phi\right] \ ,
%
%
\\
&&
T_{\alpha \beta}^{(1/2)}=\sum_A T_{\alpha \beta}^{[A]},~~{\rm with}~~T_{\alpha \beta}^{[A]} =-\frac{i}{2} 
\left[ 
    \overline{\Psi}^{[A]} \gamma_{(\alpha} \hat{D}_{\beta)} \Psi^{[A]} 
-  \left\{ \hat{D}_{(\alpha} \overline{\Psi}^{[A]} \right\} \, \gamma_{\beta)} \Psi^{[A]} 
\right]  \ ,  
\label{TD}
\\
&&
T_{\alpha\beta}^{(1)}=\frac{1}{2}
( \mathcal{F}_{\alpha \sigma }\bar{\mathcal{F}}_{\beta \gamma}
+\bar{\mathcal{F}}_{\alpha \sigma } \mathcal{F}_{\beta \gamma}
)g^{\sigma \gamma}
-\frac{1}{4}g_{\alpha\beta}\mathcal{F}_{\sigma\tau}\bar{\mathcal{F}}^{\sigma\tau}+\frac{1}{2}\mu^2\left[  
\mathcal{A}_{\alpha}\bar{\mathcal{A}}_{\beta}
+\bar{\mathcal{A}}_{\alpha}\mathcal{A}_{\beta}
-g_{\alpha\beta} \mathcal{A}_\sigma\bar{\mathcal{A}}^\sigma\right] \ .~{~~}
\label{TP}
\end{eqnarray}
The corresponding matter field equations are:
\begin{eqnarray}
&& \nabla^2 \Phi-\mu^2\Phi=0 \ , \qquad    \hat{\slashed D}\Psi^{[A]}   - \mu \Psi^{[A]}   = 0 \ , \qquad 
\nabla_\alpha\mathcal{F}^{\alpha\beta}-\mu^2 \mathcal{A}^\beta=0\ .
\label{LP2}
\end{eqnarray}
In the Proca case, 
the field eqs. (\ref{LP2}) imply the Lorentz condition, which is a dynamical requirement, rather than a gauge choice
 \cite{Brito:2015pxa,Herdeiro:2016tmi},  $\nabla_\alpha\mathcal{A}^\alpha = 0$.

In all case, the action of the matter fields, collectively denoted as $\mathcal{U}=\{ \Phi, \Psi, \mathcal{A}\}$, possesses a
  global $U(1)$ invariance, under the transformation $\mathcal{U} \rightarrow e^{i a}\mathcal{U} $, with $a$ constant.
This	implies the existence of a conserved 4-current,
which reads, respectively
\begin{eqnarray}
\label{jS}
&& j^\alpha_{(0)}=-i (\bar \Phi \partial^\alpha \Phi-\Phi \partial^\alpha \bar \Phi) \ , \qquad  j^\alpha_{(1/2)}=\bar \Psi \gamma^\alpha \Psi \ , \qquad  j^\alpha_{(1)}=
\frac{i}{2}\left[\bar{\mathcal{F}}^{\alpha\beta}\mathcal{A}_\beta-\mathcal{F}^{\alpha \beta}\bar{\mathcal{A}}_\beta\right]\ .
\end{eqnarray}
This current is conserved via the field equations, $j^{\alpha}_{ (s) ;\alpha}=0$. 
It follows that integrating the timelike component of this 4-current on a spacelike slice $\Sigma$ yields a conserved quantity --
 the \textit{Noether charge}:
\begin{eqnarray}
\label{Q}
Q_{(s)}=\int_{\Sigma}~j^t _{(s)}\ .
\end{eqnarray}
Explicit expressions for this charge will be given below for each case, $cf.$~\eqref{Qs}-\eqref{Q-S}. Upon quantization, $Q=N$, where $N$ is the particle number discussed in the Introduction.

\section{The ansatz, equations of motion and explicit physical quantities}
\label{sec3}
In this paper we shall focus on spherically symmetric configurations. The corresponding spacetime metric is most conveniently studied in Schwarzschild-like coordinates,
within the following metric ansatz:
\begin{eqnarray}
\label{metric}
 ds^2=-N(r)\sigma^2(r) dt^2+\frac{dr^2}{N(r)}+r^2 (d\theta^2+\sin^2\theta d\varphi^2) \ ,
~~~{\rm with}~~N(r)\equiv 1-\frac{2m(r)}{r}\ .
\end{eqnarray}
This ansatz introduces two radial functions: the mass function $m(r)$ and $\sigma(r)$.

In the scalar case, the matter field ansatz which is compatible with a spherically symmetric geometry is written in terms of a single real function $\phi(r)$, and reads:
\begin{eqnarray}
\label{S}
&& 
\Phi=\phi(r)e^{-iw t} \ .
\end{eqnarray}
In the Proca case, the ansatz introduces two real potentials, $F(r)$ and $G(r)$~\cite{Brito:2015pxa}:
\begin{eqnarray} 
&&
\mathcal{A}=\left[F(r)dt+iG(r)dr\right] e^{-iwt} \ .
\end{eqnarray}
In the case of a Dirac field, the ansatz also introduces two real functions, $f(r)$ and $g(r)$, but in a more cumbersome fashion. 
For a spherically symmetric configurations we have to consider two Dirac fields, $A=1,2$, 
with\footnote{Another  ansatz leading to a different set of Einstein-Dirac solutions, is also possible~\cite{Finster:1998ws}.   
However, such solutions possess very similar features and will not be considered here.}
\begin{eqnarray}
&&
\Psi^{[1]} = \begin{pmatrix} 
\cos(\frac{\theta}{2}) z(r)
\\ 
i \sin(\frac{\theta}{2}) \bar z(r)
\\ 
-i \cos(\frac{\theta}{2}) \bar z(r)
\\ 
-\sin(\frac{\theta}{2})  z(r)
\end{pmatrix}
e^{i(\frac{1}{2}\varphi-w t) } \ , \qquad
\label{down}
\Psi^{[2]} = \begin{pmatrix} 
i \sin(\frac{\theta}{2}) z(r)
\\ 
 \cos(\frac{\theta}{2}) \bar z(r)
\\ 
 \sin(\frac{\theta}{2}) \bar z(r)
\\ 
i \cos(\frac{\theta}{2})   z(r)
\end{pmatrix}
e^{i(-\frac{1}{2}\varphi-w t) } \ ,
\end{eqnarray}
where
\begin{eqnarray}
\label{z}
z(r)\equiv (1+i)f(r)+(1-i)g(r) \ .
\end{eqnarray}
For either spinor, the $individual$ energy-momentum tensor is not spherically symmetric, since
$T_\varphi^{t[A]} \sim \sin^2 \theta$, whereas the other components of $T_{\alpha}^\beta$ vanish or depend on $r$ only. However, $T_\varphi^{t[1]}+T_\varphi^{t[2]}=0$, such that the full configuration is spherically symmetric,
being compatible with the line-element (\ref{metric}).

The Einstein field equations with the energy momentum-tensors \eqref{TS}-\eqref{TP}, plus the matter field equations \eqref{LP2}, together with the ansatz \eqref{S}-\eqref{down}, lead to a system of three (four) coupled ordinary differential equations for the the scalar (Dirac and Proca) cases.  The equation for the mass function $m(r)$ reads
\begin{equation}
m'=4 \pi G r^2 \mathcal{X}_{(s)} \ ,
\end{equation}
where
\begin{equation}
\label{m-S}
\mathcal{X}_{(0)}=
N \phi'^2+\mu^2 \phi^2+\frac{w^2 \phi^2}{N\sigma^2} \ , \qquad  \mathcal{X}_{(1/2)}=\frac{8w(f^2+g^2)}{\sqrt{N}\sigma}  \ , \qquad \mathcal{X}_{(1)}=\frac{(F'-wG)^2}{2\sigma^2}
+\frac{\mu^2}{2} \left(G^2N+\frac{F^2}{N\sigma^2}\right) \ . 
\end{equation}
Using the Dirac equation, $\mathcal{X}_{(1/2)}$ can exhibit a structure more similar to its bosonic counterparts:
\begin{eqnarray}
\label{m-D1}
  \mathcal{X}_{(1/2)}=8 
\left[
\sqrt{N}(gf'-fg')
+\frac{2fg}{r}
+\mu(g^2-f^2)
\right]\ .
\end{eqnarray}

The equation for the metric function $\sigma(r)$ reads
\begin{equation}
\frac{\sigma'}{\sigma}=4 \pi G r \mathcal{Y}_{(s)} \ ,
\end{equation}
where
\begin{equation}
\label{s-S}
\mathcal{Y}_{(0)}=2
\left( 
\phi'^2+\frac{w^2\phi^2}{N^2\sigma^2}
\right) \ , \qquad \mathcal{Y}_{(1/2)}=\frac{8}{\sqrt{N}}
\left[
gf'-fg'+\frac{w(f^2+g^2)}{N\sigma}
\right] \ , \qquad \mathcal{Y}_{(1)}=\mu^2
\left(G^2+\frac{F^2}{N^2\sigma^2} \right)\ .
\end{equation}
Finally, the equations for the functions in the matter fields, $\phi$ (scalar), $f,g$ (Dirac) and $F,G$ (Proca) 
are\footnote{Observe~\eqref{fgeq} are actually two equations: the upper $f$ and lower $g$ equations. }
\begin{eqnarray}
\label{e-S}
&&
\phi''+\left(\frac{2}{r}+\frac{N'}{N}+\frac{\sigma'}{\sigma}\right)\phi'
+
\left(\frac{w^2}{N\sigma^2}-\mu^2\right)\frac{\phi}{N}=0\ ,
\\
\label{fgeq}
&&
\begin{pmatrix} 
f' \\ g'
\end{pmatrix} 
+
\left(
\frac{N'}{4N}+\frac{\sigma'}{2\sigma}\pm\frac{1}{r\sqrt{N}}+\frac{1}{r} 
\right)
\begin{pmatrix} 
f \\
g
\end{pmatrix} 
+
\left(
\frac{\mu}{\sqrt{N}}\mp \frac{w}{ N\sigma}
\right)
\begin{pmatrix} 
g \\
f
\end{pmatrix} 
=0\ ,
\\
&&
 \frac{d}{dr}\left\{\frac{r^2[F'-wG]}{\sigma}\right\}=\frac{\mu^2r^2F}{\sigma N} \ , \qquad  wG-F'=\frac{\mu^2\sigma^2 N G }{w} \ .
\end{eqnarray}
For each case, there is a supplementary  second order constraint equation 
between the metric functions $m(r)$ and $\sigma(r)$,
which, however, is a differential consequence of the above field equations.
 
Let us also provide explicit expressions for two relevant physical quantities, in terms of the ansatz \eqref{S}-\eqref{down}. The energy density measured by a static observer, $\rho=-T_t^t$, is, from \eqref{TS}-\eqref{TP},
\begin{equation}
\label{rho-S}
\rho_{(0)}=N\phi'^2+\left(\mu^2+\frac{w^2}{N\sigma^2}\right)\phi^2\ , \qquad
\rho_{(1/2)}=\frac{8w(f^2+g^2)}{\sqrt{N}\sigma}\ , \qquad
\rho_{(1)}= \frac{(F'-wG)^2}{2\sigma^2}
+\frac{1}{2}\mu^2 \left(G^2N+\frac{F^2}{N\sigma^2} \right) \ .
\end{equation}
Also, the Noether charge, computed from~\eqref{Q} is\footnote{For the Einstein-Dirac system,
(\ref{Qs}) corresponds to a single spinor,
 the total Noether charge for the solutions here being $Q=2Q_{(1/2)}$.}
\begin{equation}
\label{Qs}
Q_{(s)}= 8\pi \int^{\infty}_0 dr\, r^2 \mathcal{Z}_{(s)} \ ,
\end{equation}
where
\begin{equation}
\label{Q-S}
\mathcal{Z}_{(0)}= w  \frac{\phi^2}{N\sigma}\ , \qquad \mathcal{Z}_{(1/2)}= 2  \frac{(f^2+g^2)}{\sqrt{N}}\ , \qquad
\mathcal{Z}_{(1)}=  \frac{(wG-F')G}{\sigma} \ .
\end{equation}

\subsection{Units and scaling symmetries}

For guidance, let us briefly comment on the physical dimensions of the fundamental fields in (\ref{action}), using $L=$~Lenght:
\begin{equation}
[\Phi]=\frac{1}{L}\ , \qquad [\Psi]=\frac{1}{L^{3/2}}\ , \qquad [\mathcal{A}_\alpha] =\frac{1}{L}\ .
\end{equation}
In all three cases,
the factor of $4 \pi G$ in the Einstein field equations can be set to one by a redefinition of 
 the matter functions 
\begin{equation}
\{\phi,f,g,F,G\} =\frac{1}{\sqrt{4\pi G}} \{\bar \phi, \bar f, \bar g, \bar F, \bar G \} \ ,~
\label{t-D}
\end{equation}
Since $[G]=L^2$, the scaled functions $\bar \phi,\bar F$,  $\bar G$
are dimensionless, while $[\bar f]=1/\sqrt{L}$, $[\bar g]=1/\sqrt{L}$.


In practice, the solving of the equations of motion makes use of some scaling invariances thereof. We observe that
in the scalar and Dirac cases,
the equations of motion possess the  symmetry
\begin{eqnarray}
\label{s0}
(s0):~~~\{\sigma,w\} \to \lambda \{\sigma,w\}\ ,
\end{eqnarray}
with $\lambda$ a positive constant. In the Proca case~\eqref{s0} together with $F\rightarrow \lambda F$ Êis the corresponding symmetry. 
Another, more central, invariance holds, which, however, acts differently on the matter field variables, depending on the spin of the field. It reads
\begin{equation}
\label{sca1}
(s1):~~~
\{ r, m  \}= \lambda \{\bar r,\bar m \} \ , \ \ \  \{w,\mu\}=\frac{1}{\lambda}\{\bar w,\bar \mu \}\ ,   \ \ \  \sigma =\bar \sigma  \ , \ \ \ 
\left\{
\begin{array}{l} 
\displaystyle{\phi=\bar \phi } \ , \\
\displaystyle{\{f , g  \}=\frac{1}{\sqrt{\lambda}}\{ \bar f , \bar g  \}} \ , \\
\displaystyle{\{F , G  \}=\{ \bar F , \bar G  \}} \ .
\end{array} 
\right.
\end{equation}
In all cases the product $m(r)\mu$ is left invariant by the symmetry $(s1)$.
This is also the case for the ratio $w/\mu$.
The $(s1)$ invariance is usually used to work in units set by the field mass,
\begin{eqnarray}
\bar \mu=1\ , ~~i.e.~~\lambda=\frac{1}{\mu}\ .
\label{mbar}
\end{eqnarray}
Since $[\mu]=1/L$,~\eqref{mbar} together with (\ref{t-D}), leads also to dimensionless spinor functions $\bar f$ and $\bar g$.
Then, the full scalings to obtain dimensioness spinor functions read:
 \begin{eqnarray}
\{f,g\}=\frac{\sqrt{\mu}}{\sqrt{4\pi G}}\{\bar f,\bar g\} \ .
\end{eqnarray}

To summarize, in practice and for all cases, the field equations are solved in units which amount to set
$4\pi G=1$, $\mu=1$, in the equations of motion. Then, we use a numerical shooting method (detailed below) with the (only) input parameter:
\begin{eqnarray}
\bar w=\frac{w}{\mu}\ .
\end{eqnarray}
As such, the physical mass of a solution, the ADM mass, denoted as $M$, is related to the mass obtained from the numerical procedure, $M_{(num)}$, by $\mu$ and $G$, as ($M_{Pl}={1}/{\sqrt{G}}$ denotes the Planck mass)
\begin{eqnarray}
M_{(num)}=\frac{\mu M}{M_{Pl}^2}\  .
\end{eqnarray}
Similarly, the Noether charge of the solutions relates to the one obtained from the numerical procedure, $Q_{(num)}$ (performed in units
with $4\pi G=1$, $\mu=1$) as, in all three cases,
 \begin{eqnarray}
\label{Qn}
Q=\frac{1}{4 \pi G} \frac{Q_{(num)}}{\mu^2} \ .
\end{eqnarray}


The effect of the scaling symmetry $(s1)$
on the global mass and charge is
\begin{eqnarray}
M_{(num)}= \lambda \bar M_{(num)}\ ,~~Q_{(num)}= \lambda^2 \bar Q_{(num)} \ .
\label{sca2}
\end{eqnarray}
Let us suppose we have a numerical solution with some values
($M_{(num)}$, $Q_{(num)}$).
Then, we can use the above symmetry in order to obtain a solution with
\begin{eqnarray}
 \bar Q_{(num)}=Q_0\ .
\end{eqnarray}
This amounts to fixing the value of the scaling parameter $\lambda$,
\begin{eqnarray}
 \lambda=\sqrt{\frac{Q_{(num)}}{Q_0}}\ ,
\end{eqnarray}
such that the $numerical$ mass of the new solution will be
\begin{eqnarray}
 \bar M_{(num)}= M_{(num)}  \sqrt{\frac{Q_0}{ Q_{(num)}}} \ .
\end{eqnarray}
The values of $w$ and $\mu$
should also be scaled accordingly, as given by $(s1)$.
For example, taking $Q_0=1$ normalizes the total charge of a fermion to unity.\footnote{The $physical$
total charge, however, is not $\bar Q_{(num)}$, but rather (\ref{Qn}).}

\section{The solutions}
\label{sec4}

\subsection{The boundary conditions and numerical method}
The solutions to the three/four coupled ordinary differential equations presented in the previous section is obtained numerically, after imposing suitable boundary conditions, as we now describe.

The boundary conditions satisfied by the  metric functions at the origin are
\begin{eqnarray}
m(0)=0\ ,~~\sigma(0)=\sigma_0\ ,
\end{eqnarray}
while at infinity  one imposes
\begin{eqnarray}
m(\infty)=M,~~\sigma(\infty)=1\ ,
\end{eqnarray}
with  $\sigma(0)$, $M$  numbers fixed by numerics.
The above conditions for $\sigma$
fixes the symmetry (\ref{s0}) of the system.
The boundary conditions satisfied by the matter functions 
are more involved.
First, in all cases, they should vanish as $r\to \infty$
\begin{eqnarray}
\phi(\infty)=f(\infty)=g(\infty)=F(\infty)=G(\infty)=0\ .
\end{eqnarray}
This is basically due of the presence of the mass term in the action and the requirement of asymptotic flatness.
Additionally, an analysis of the field equations 
near the origin leads to the following boundary conditions for the matter functions 
\begin{equation}
\frac{d\phi(r)}{dr}\bigg|_{r=0}=0\ , \qquad f(0)=0,~~\frac{dg(r)}{dr}\bigg|_{r=0}=0\ , \qquad \frac{dF(r)}{dr}\bigg|_{r=0}=0,~~G(0)=0\ . 
\end{equation}
In each case, one can construct 
an approximate form of the solutions, compatible with the boundary conditions above 
(see $e.g$~\cite{Brito:2015pxa} for the Proca case).

The numerical construction of the full solutions is straightforward.
In all cases, we use a standard Runge-Kutta ordinary differential equation solver and evaluate
the initial conditions at $r=10^{-6}$ for global tolerance $10^{-14}$,
 adjusting for fixed shooting
parameters (which are some constants which enter the near origin expression of the solutions)
and integrating towards $r\to \infty$.
The accuracy of the solutions was also monitored by computing virial relations satisfied by these systems.
 For a given $w$,
the solution form a discrete set indexed by the number of nodes, $n$,
of (some of) the matter function(s).
The data shown in this work correspond to fundamental solutions,
except for Fig.~\ref{fig4} (right panel).

\subsection{Numerical results: domain of existence and some properties}
The solutions obtained are, in all cases, topologically trivial, with $0\leqslant r<\infty$. They possess no horizon, while the size of  the $S^2$-sector of the metric shrinks to zero as $r\to 0$, $cf.$~\eqref{metric}. The latter limit is just the standard coordinate singularity of spherical coordinates. Indeed, the solutions are everywhere regular and asymptotically flat. 

The domain of existence of the solutions, in all three cases, corresponds to a spiral in an ADM mass $M$, $vs.$  frequency, $w$, diagram, starting from $M=0$ for $w=\mu$, in which limit the fields becomes very diluted and the solution trivialises - Fig.~\ref{fig2} (left panel).  At some intermediate frequency, a maximal mass is attained. These  masses and corresponding frequencies are given in the second and third columns of Table 1.  As a trend, one can see that the maximal mass increases with the spin. In each case there is also a minimal frequency, below which no solutions are found. As can be seen in Table 1, for the minimal frequency the behaviour is not monotonic with spin. After reaching the minimal frequency, the spiral backbends into a second branch. In all three cases we were able to obtain further backbendings and branches. The frequencies of the first (minimal frequency), second and third backbendings are shown in the 4$^{th}$-6$^{th}$ columns of Table~1.  Likely, these spirals approach, at their centre, a critical singular solution. We have not, however, been able to follow the spiral that far. 

\begin{table}[h!]
{\small \begin{center}
\begin{tabular}{c|||c|c||c|c|c||c|c}
\hline
& \ \ $M^{\rm max}$ & \ \ $w(M^{\rm max})$ & \ \  $w^{\rm 1st}$ & \ \ $w^{\rm 2nd}$ & \ \ $w^{\rm 3rd}$ & $M=Q$ & \ \ $w^{\rm crossing}$\\
\hline
scalar \ \  & \ \   0.633 \ \ & \ \     0.853 \ \ &  \ \  0.768 \ \ & \ \ 0.856  \ \ & \ \ 0.840 \ \ & \ \ 0.552 & \ \ 0.778 \\
Dirac  \ \  & \ \   0.709 \ \ & \ \     0.830 \ \  & \ \  0.733 \ \ & \ \ 0.859  \ \ & \ \ 0.824 \ \ & \ \ 0.710 & \ \ 0.823  \\
Proca  \ \  & \ \   1.058 \ \ & \ \     0.875 \ \  & \ \  0.814 \ \ & \ \ 0.892  \ \ & \ \ 0.891 \ \ & \ \ 0.905 & \ \ 0.818 \\
 \hline
\end{tabular}\\
\bigskip
 \caption{1$^{st}$ column: the three different models. 2$^{nd}$ column: maximal mass of fundamental solutions. 3$^{rd}$-6$^{th}$ columns: frequencies of maximal mass solution and at the first (minimum frequency), second and third backbending in the Mass $vs.$ frequency diagram of Fig.~\ref{fig2} (left panel). 7$^{th}$-8$^{th}$ columns: parameters of the solutions with equal ADM mass and Noether charge. All quantities are presented in units of $\mu$, $G$. }
\end{center}}
\end{table}


\begin{figure}[h!]
\begin{center}
\includegraphics[width=0.49\textwidth]{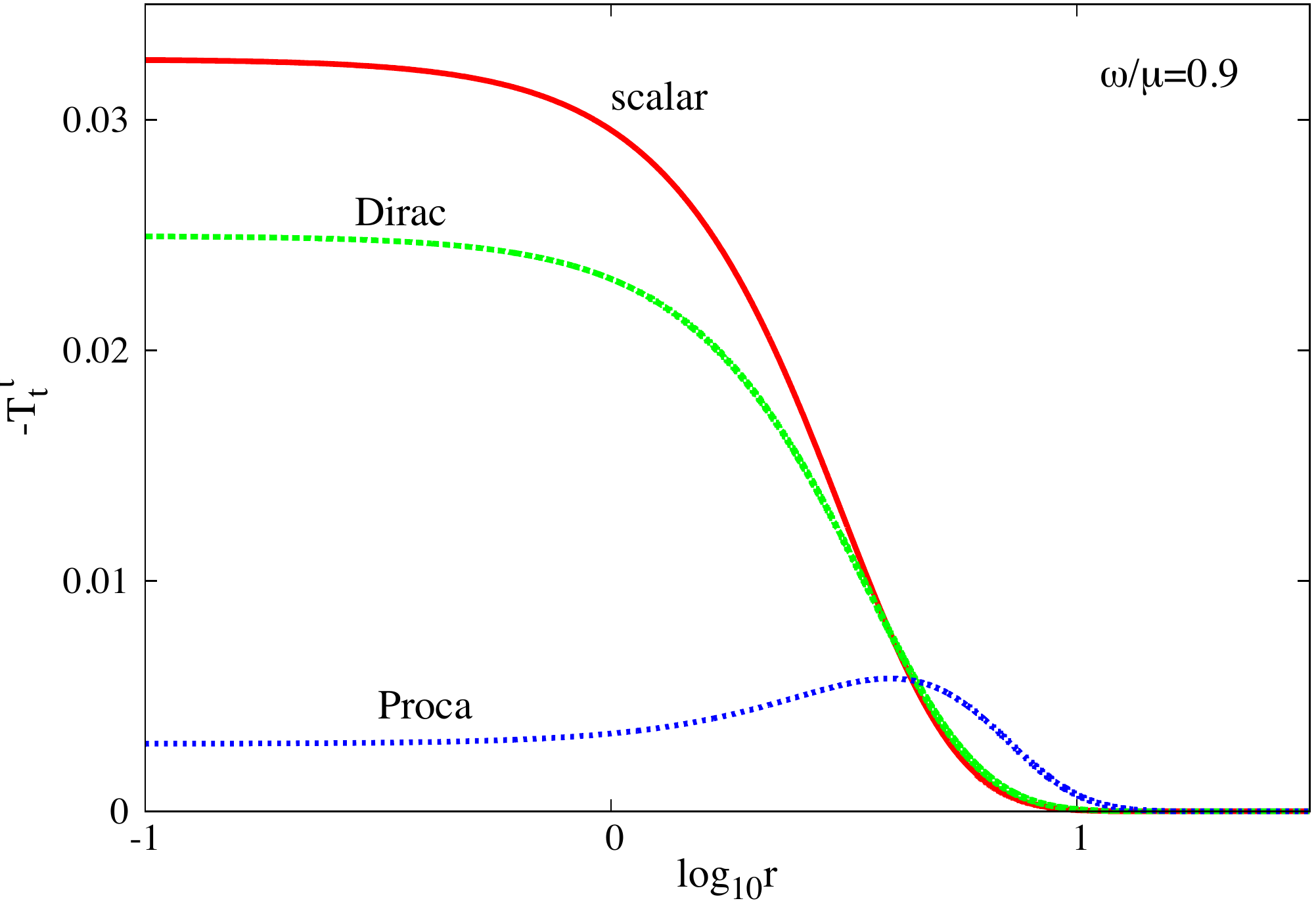}\ \ \
\includegraphics[width=0.49\textwidth]{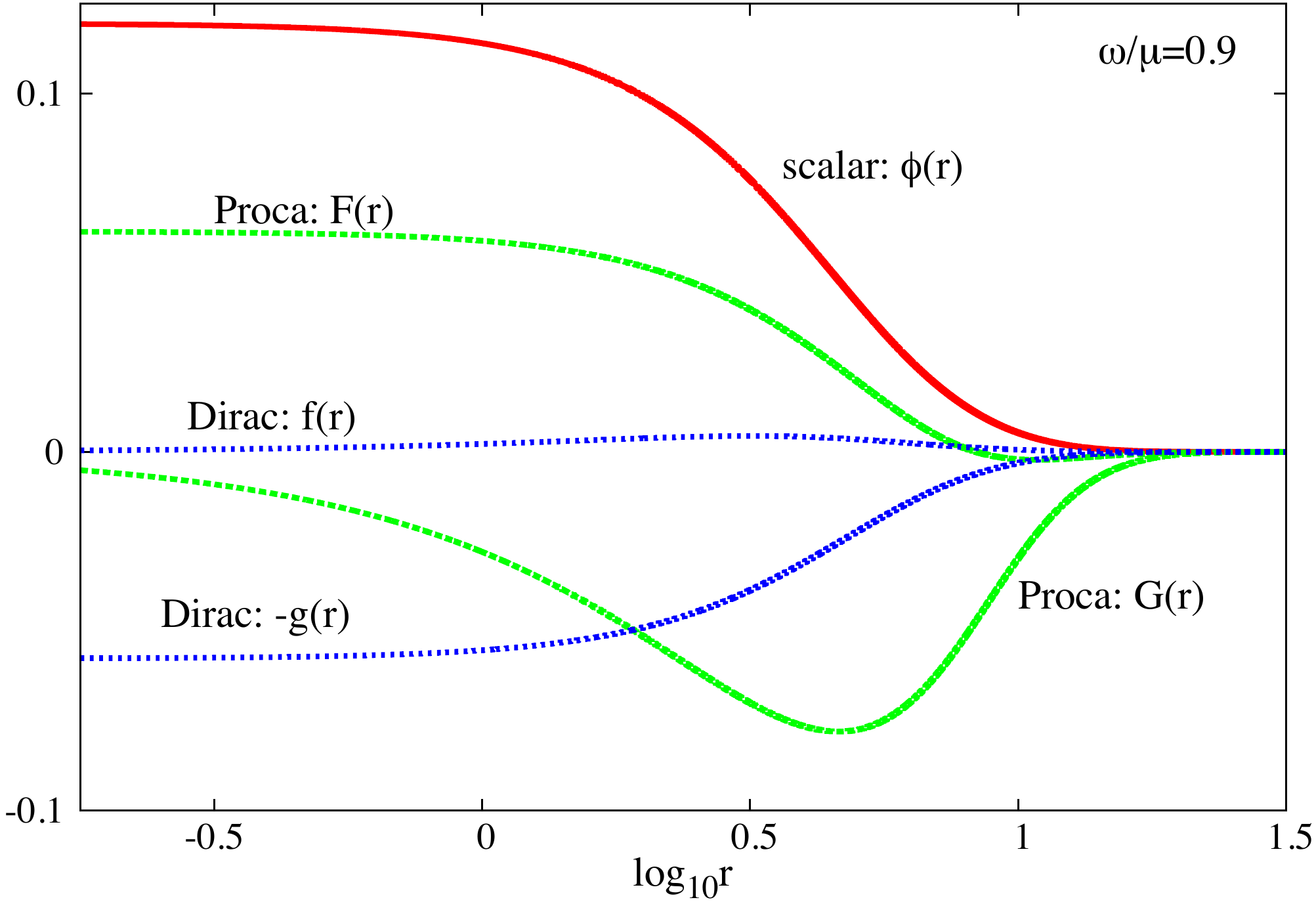}
\includegraphics[width=0.49\textwidth]{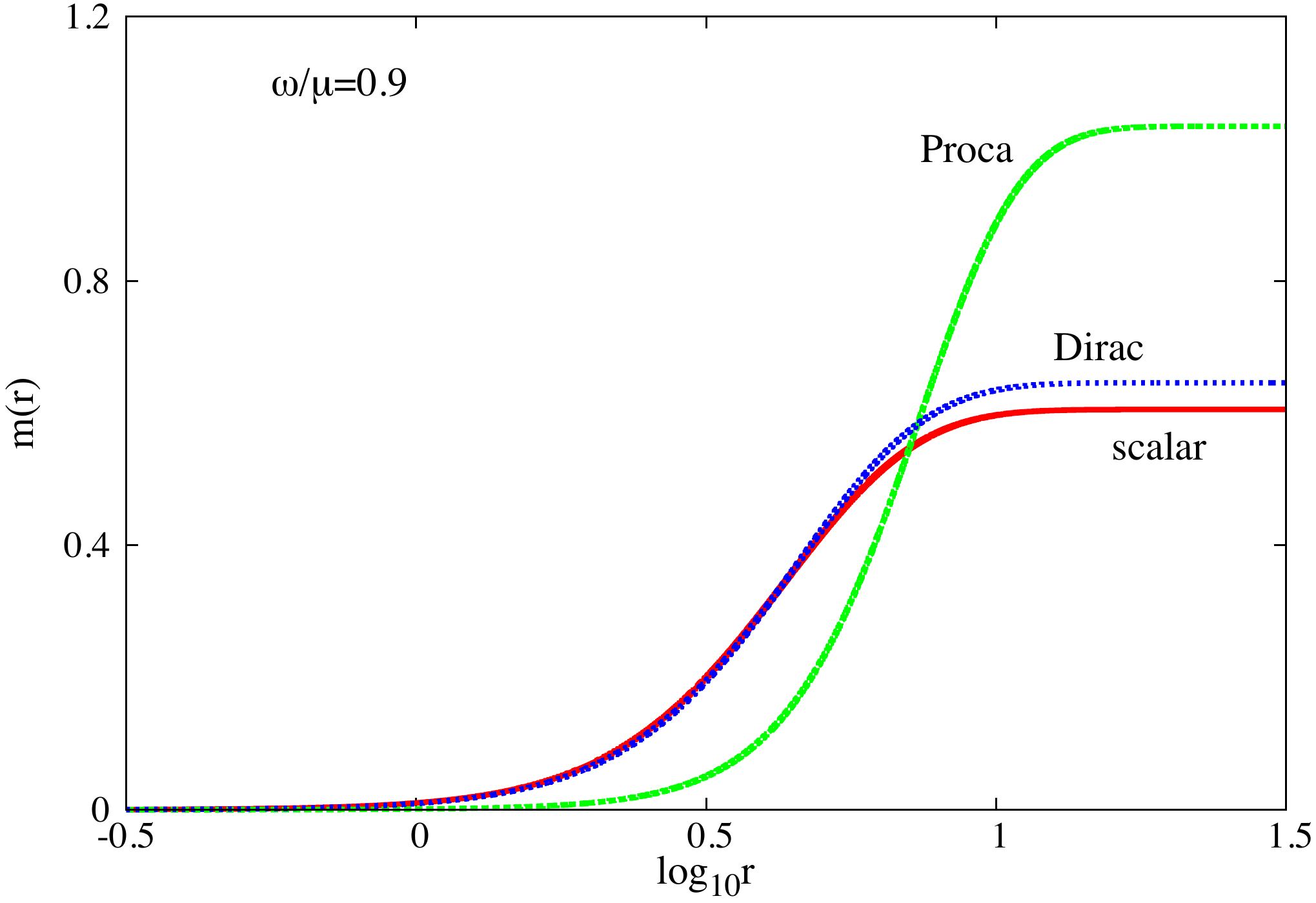}\ 
\includegraphics[width=0.49\textwidth]{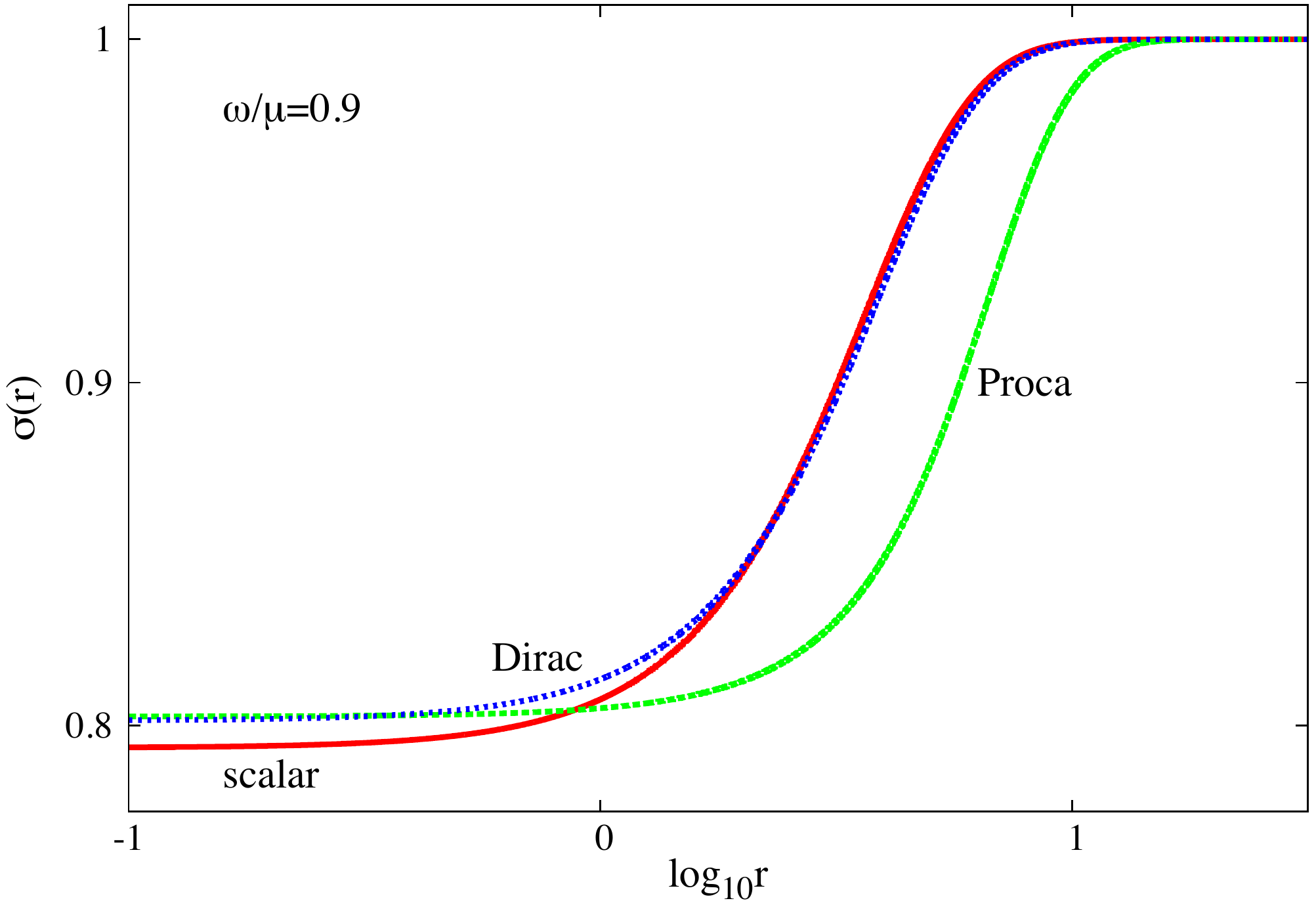}
\caption{\small{Top left panel: the energy density, defined by eqs.~\eqref{rho-S}, is shown for a (typical) solution of each model, all with the same frequency to particle mass ratio, $w/\mu=0.9$. Top right panel: the matter functions profiles for the same solutions. Bottom panels: the metric functions profiles for the same solutions. }}
\label{figprofile}
\end{center}
\end{figure}  

As already mentioned, the Noether charge is a measure of the particle number. As such, it can also give us a criterion for stability. If the Noether charge multiplied by the quanta mass $\mu$ is smaller than the ADM mass $M$, then the solution has excess energy and it should be unstable against fission. In all three cases we confirmed that close to the maximal frequency, $w=\mu$ the solutions have a Noether charge larger than the ADM mass (in units of $\mu$). This corresponds to a regime where there is binding energy, a necessary, albeit not sufficient,  condition for stability. At some point, however, the Noether charge and ADM mass curves cross and $M$ becomes larger than $Q$ corresponding to solutions with excess energy and hence unstable. The crossing frequency, and corresponding $M=Q$ are given in the 7$^{th}$-8$^{th}$  columns of Table 1. Whereas this energy analysis is meaningful in the bosonic case, in the fermionic case it is not so. We will come back to this point below.

Within the solutions with binding energy not all are stable. Stability has been established for the solutions that exists between the flat space limit $w=\mu$ and the maximal mass for both the scalar~\cite{Gleiser:1988ih,Lee:1988av} and Proca cases~\cite{Brito:2015pxa,Sanchis-Gual:2017bhw}. It is reasonable to expect the same holds for the Dirac case.\footnote{The results in \cite{Finster:1998ws}
show the existence of stable solutions in Einstein-Dirac model, although after imposing the single particle condition.
}

In Fig.~\ref{figprofile} we exhibit the energy density profiles (top left panel), the matter-functions profiles (top right panel) and the metric functions profiles (bottom panels) for illustrative solutions in all three cases. As can be seen, all of them correspond to localised lumps of energy. Curiously, the Dirac profile mimics closely the scalar one, whereas the Proca one is qualitatively different, exhibiting the maximum for the energy density away from the origin. It is also interesting to observe that for the Proca case, the matter function $F(r)$ 
necessarily exhibit one node 
\cite{Brito:2015pxa}. 
Somewhat unexpectedly, for the rotating Proca solutions, 
the corresponding function is nodeless for fundamental states~\cite{Herdeiro:2017phl}.

\subsection{Bosonic $vs.$ fermionic solutions}

If we impose\footnote{Actually, in the Dirac  case we impose $Q=1$ for each spinor. 
}
$Q=1$ -- which is a mandatory requirement for Dirac stars, but optional for boson and Proca stars --, 
the spiral in Fig.~\ref{fig2} (left panel)
is not a sequence of solutions with fixed $\mu$ and varying $Q$; it is a sequence with fixed $Q$ and varying $\mu$. Consequently, one is discussing \textit{a sequence of solution of different models} ($\mu$ is a parameter in the action). 

\bigskip

\begin{figure}[h!]
\begin{center}
\includegraphics[width=0.495\textwidth]{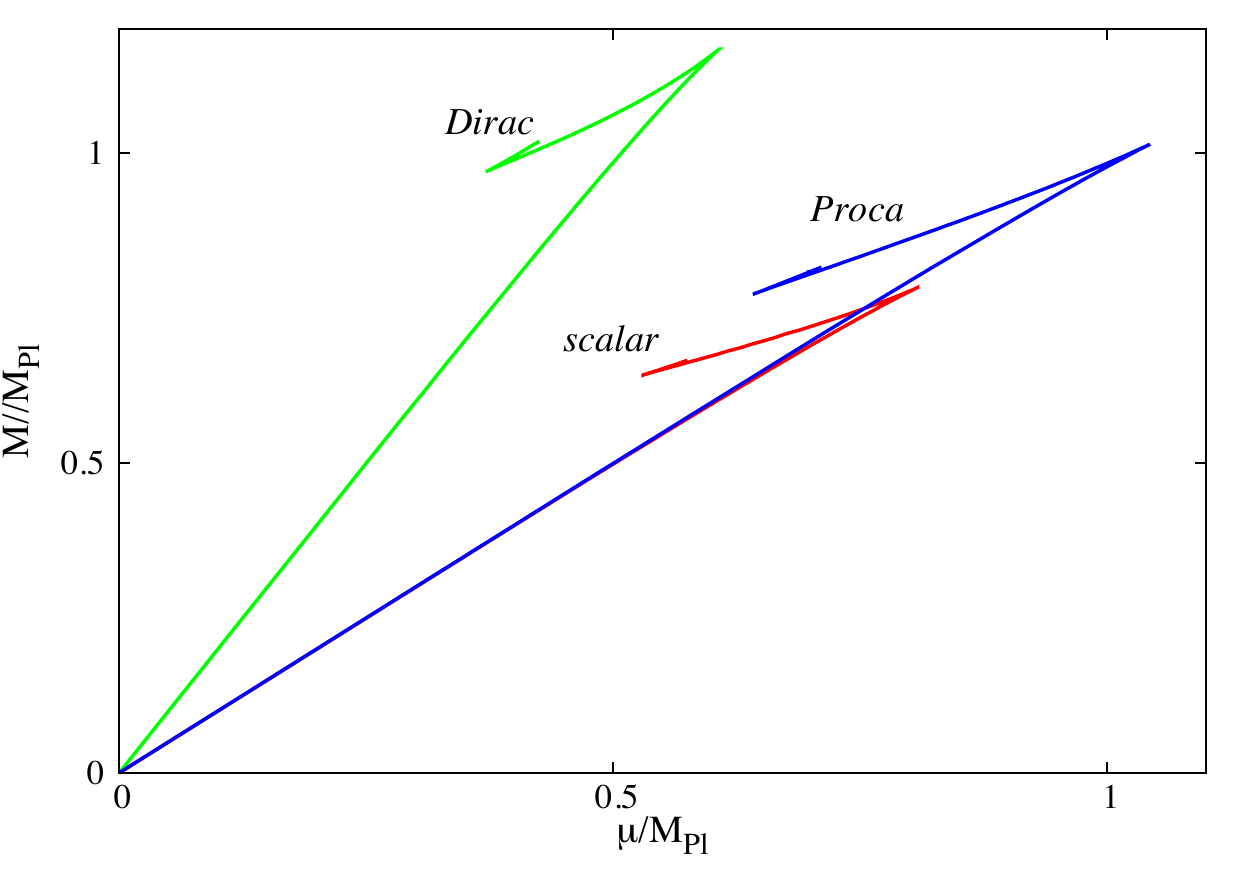}
\includegraphics[width=0.495\textwidth]{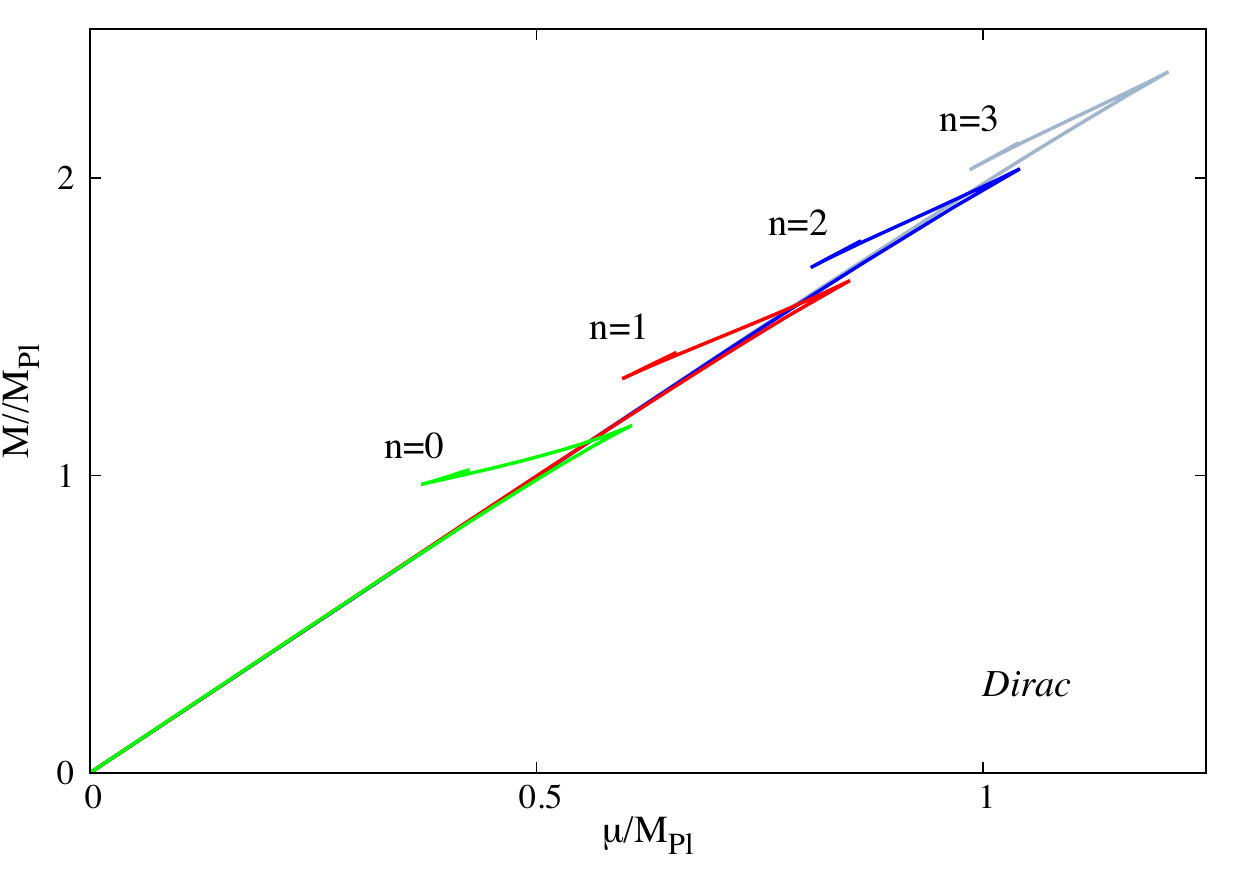} 
\caption{\small{(Left panel) ADM mass $vs.$ scalar field mass, in Planck units, for the three families of stars of fundamental fields. (Right panel) Same for the fundamental ($n=0$) and the first three ($n=1,2,3$) excited states of the Dirac field.
The single particle condition, $Q=1$, is imposed here.
}}
\label{fig4}
\end{center}
\end{figure}

An immediate (and expectable) consequence is that there cannot be a discrepancy of orders of magnitude between the physical mass of the configuration, $M$, and the mass of the quanta, $\mu$. Indeed, they should be roughly \textit{of the same order of magnitude}, in sharp contrast to the macroscopic quantum states described above. This is precisely what one may appreciate in Fig.~\ref{fig4} (left), where we plot the same data as in Fig.~\ref{fig2} but imposing the single particle condition.\footnote{As discussed above, this condition may be imposed in a straightforward manner by using the  scaling symmetry of the solutions, \eqref{sca1} and \eqref{sca2}. 
} 

\bigskip

A novel consequence of treating the solutions as one particle \textit{microscopic classical configurations} 
is that not only the total mass, $M$, is bounded, but the mass of the field $\mu$ is \textit{also bounded}, and, for fundamental states, never exceeds, roughly, $M_{Pl}$.  Thus, the aforementioned intuitive bound -- that self-gravitating standing waves cannot exceed a certain total energy --, translates for these single particle configurations into the requirement that the particle's size (measured by its Compton wavelength) cannot be smaller than a certain size ($\sim$ Planck length). 
This upper $\mu$ bound can be pushed further up by considering excited configurations which are indexed by $n$, the node number of the field amplitude(s), making these configurations increasingly trans-Planckian. 
The corresponding masses for the Dirac model are shown in Fig.~\ref{fig4}. (right). 
Interestingly, for  Dirac fields, the $n\geqslant 1$ 
excited states are not necessarily~\cite{Finster:1998ws}  unstable (unlike bosons~\cite{Lee:1988av}).

\section{Conclusions and remarks}
\label{sec5}

The main purpose of this work was to provide a comparative analysis 
of three different types of solitonic solutions of GR-matter systems,
which can be interpreted as explicit realizations of Wheeler's geon concept for matter fields of spin
$0$, $1$ and $1/2$, respectively.
As classical field theory solutions, our results show that
the existence of these self-gravitating, stable, energy lumps, composed of standing waves, 
does not distinguish between the fermionic/bosonic nature of the field, possessing a variety of  similar features. 
However, if one imposes that the configuration describes a single particle, as required for fermions, 
one finds that for each field mass there is a discrete set of  
 states, up to a maximal field mass.  

Since geon-inspired solutions exist in classical field theories of spins $0,1/2,1$, it is  
likely they may exist for \textit{any} spin, given a consistent matter model minimally coupled to GR, likely with similar properties.
As the simplest extension, 
we predict the existence of fermion stars with spin $3/2$ fields, 
which satisfy the conditions ${\bf (i)}$-${\bf (iii)}$ in the Introduction.
 
\medskip
In this context,
it is interesting to mention that the observed similarities between bosonic and fermionic 
solitons also hold in the absence of gravity.
The extra-interaction necessary for the existence of a localized solution is provided by turning on nonlinear terms in the Lagrangian.
This results in (flat space) $Q$-ball type solutions 
discussed $e.g.$
in 
\cite{Coleman:1985ki,Soler:1970xp,Loginov:2015rya}
for spin $0$, $1/2$ and $1$, respectively.
Moreover, one can show that these configurations allow for rotating generalizations.
In fact, configurations with a nonvanishing angular momentum 
exist as well in the (gravitating) model (\ref{action}) -- see \cite{Schunck:1996he,Yoshida:1997qf}
for spinning Klein-Gordon
 and Proca \cite{Brito:2015pxa} geons;  the results for the Dirac case will be reported 
elsewhere.

On the other hand, a striking \textit{difference} between bosonic and fermionic solutions still seems to exists: 
while boson stars allow for a  black hole horizon to be placed inside them (provided the full configuration
is rotating subject to a synchronisation condition  
\cite{Herdeiro:2014goa,Herdeiro:2016tmi}),
no such configurations are known for a Dirac field.
Usually, this is viewed as a consequence of the absence of 
superradiance for a fermionic field on the Kerr background \cite{Brito:2015oca}.
But spinning black holes 
with scalar hair 
may exist even in the absence of the superradiant instability, the hair being intrinsically non-linear
\cite{Brihaye:2014nba,Herdeiro:2015kha}.
Therefore one cannot \textit{a priori} exclude that a fermion field also shares this feature, and as a consequence,
 Dirac stars could
allow for black hole generalizations provided they rotate synchronously
with the horizon.
  
\medskip
Finally, returning to a more fundamental level,
in Wheeler's view, the concept of geon ``\textit{completes the scheme of classical
physics by providing for the first time an acceptable
classical theory of the concept of body.}" [pag. 536], but ``\textit{one's interest in following geons into
quantum domain will depend upon one's view of the relation
between very small geons and elementary particles.}" [pag 512].
The prevailing view, at present, is that the classical GR geometric picture is inadequate for  the quantum world, 
where quantum fluctuations are of the order of the spacetime metric.\footnote{
One could also object that the fermionic field is not quantized in the treatment discussed here. In this respect it is amusing to recall the following discussion~(\cite{book}, p. 143) between de Witt and Wheeler, concerning~\cite{Brill:1957fx}:\\ 
DE WITT: \textit{``In this work the neutrinos are not quantized; they are not real
neutrinos, are they?"}\\
WHEELER: \textit{``One puts into each neutrino state just one neutrino; this
includes all of the results of second quantization."}
}  Moreover, there are no doubts that a successful description of elementary particles has been provided by Quantum Field Theory. But one may not exclude that a more conceptually fulfilling, likely complementary, description  of a single particle is possible. In this respect, geons may still have a role to play in bridging the classical and quantum world. A more conservative perspective, however, is that the (macroscopic) bosonic scalar or Proca geons herein may play a role in Nature, whereas the (necessarily microscopic) Dirac stars are
a mere mathematical exercise.

\medskip

\section*{Acknowledgements}
We would like to thank M. Zilh\~ao for the images in Figure 1. C. H. and E. R. acknowledge funding from the FCT-IF programme. This work was supported by the projects H2020-MSCA-RISE-2015 Grant  StronGrHEP-690904, H2020-MSCA-RISE-2017 Grant  FunFiCO-777740, and UID/MAT/04106/2013 (CIDMA).  The authors would also like to acknowledge networking support by the COST Action CA16104.



\end{document}